\documentclass[aps,showkeys,superscriptaddress,nofootinbib]{revtex4}
\RequirePackage[colorlinks,hyperindex]{hyperref}
\RequirePackage[english]{babel}
\RequirePackage[latin1]{inputenc}
\RequirePackage[T1]{fontenc}
\usepackage{graphicx}
\RequirePackage{mathrsfs}
\RequirePackage{amsmath}
\RequirePackage{amssymb}
\RequirePackage{amsbsy}
\RequirePackage{color}
\RequirePackage{bm}
\newcommand{\be}{\begin{equation}}
\newcommand{\ee}{\end{equation}}
\newcommand{\bea}{\begin{eqnarray}}
\newcommand{\eea}{\end{eqnarray}}
\def\de#1/de#2{\frac{\partial {#1}}{\partial {#2}}}
\newtheorem{Proposition}{Proposition}[section]
\hypersetup{colorlinks=true,breaklinks=true,urlcolor=blue,linkcolor=red}
\begin{document}
\title{A note on the junction conditions in $f({\cal Q})$-gravity}
\author{Stefano Vignolo\footnote{corresponding author}}
\email{stefano.vignolo@unige.it}
\affiliation{DIME, Universit\`{a} di Genova, Via all'Opera Pia 15, 16145 Genova, Italy}
\affiliation{INFN Sezione di Genova, Via Dodecaneso 33, 16146 Genova, Italy}
\affiliation{GNFM, Istituto Nazionale di Alta Matematica, P.le Aldo Moro 5, 00185 Roma, ITALY}
\author{Fabrizio Esposito}
\email{fabrizio.esposito01@edu.unige.it}
\affiliation{DIME, Universit\`{a} di Genova, Via all'Opera Pia 15, 16145 Genova, Italy}
\affiliation{INFN Sezione di Genova, Via Dodecaneso 33, 16146 Genova, Italy}
\author{Sante Carloni}
\email{sante.carloni@unige.it}
\affiliation{DIME, Universit\`{a} di Genova, Via all'Opera Pia 15, 16145 Genova, Italy}
\affiliation{INFN Sezione di Genova, Via Dodecaneso 33, 16146 Genova, Italy}
\affiliation{Institute of Theoretical Physics, Faculty of Mathematics and Physics,
Charles University, Prague, V Hole{\v s}ovi{\v c}k{\' a}ch 2, 180 00 Prague 8, Czech Republic}
\date{\today}
\begin{abstract}
Using the notion of distribution-valued tensor, we discuss the junction conditions within the framework of $f({\cal Q})$-gravity. We obtain the necessary and sufficient conditions for two distinct solutions of the field equations to be smoothly joined on a given separation hypersurface.
\end{abstract}
\maketitle
\section{Introduction}
General Relativity (GR) is undoubtedly a cornerstone of modern physics. However, GR still presents some shortcomings and inconsistencies at the astrophysical and cosmological scales, as well as at the quantum level. In this regard, the need to explain phenomena such as dark matter and dark energy, or to successfully integrate GR with quantum physics, has motivated the study of several alternative theories of gravitation, nowadays known as Extended Theories of Gravity or Modified Gravity (ETG, MG) \cite{CAPOZZIELLO2011167,Clifton_2012}. 

Among the different ETGs developed during the last few decades, the so-called $f({\cal Q})$-gravity has recently attracted the interest of many researchers. $f({\cal Q})$-gravity fits into the general context of metric-affine theories of gravity, where the metric and the affine connection are independent geometrical quantities. More specifically, in $f({\cal Q})$-gravity, the connection is assumed to be torsionless and flat but not metric--compatible. The gravitational Lagrangian is a function $f({\cal Q})$ of the nonmetricity scalar $\cal Q$. Essentially, $f({\cal Q})$-gravity generalizes Symmetric Teleparallel Gravity (STEGR) \cite{Nester:1998mp,Adak:2005cd,BeltranJimenez:2017tkd} just as $f(R)$ theories do compared to GR \cite{Faraoni,Defelice}. $f({\cal Q})$-gravity has been thoroughly investigated with interesting results in cosmological and astrophysical frameworks \cite{Frusciante1,Frusciante2,Vignolo:2021frk,Mimoso,Esposito:2021ect,Esposito:2022omp,Esposito:2023dgm,Khyllep:2022spx,Anagnostopoulos:2022gej,Capozziello:2023vne,Paliathanasis:2023nkb,Dimakis:2023uib,Yang:2021fjy,Dimakis:2022rkd,Calza:2022mwt,Hohmann:2021ast,DAmbrosio:2021zpm,Bahamonde:2022esv,Heisenberg:2023lru}.

In this note, we discuss the junction conditions arising from $f(\mathcal{Q})$-gravity. To the best of the authors' knowledge, this topic has only been partly addressed in the literature \cite{Esposito:2023dgm,Tayde:2023pjh,ZeeshanGul:2024qjc,Mohanty}, with non-exhaustive results. The junction conditions ensure the smooth joining of two different solutions of the field equations on a given separation hypersurface. Clearly, their broadest use concerns astrophysical applications. 

Following a well-known and consolidated approach \cite{Israel,L1,L2,Choquet,CJSClarke_1987,Barrabes:1997kk,Poisson,Vignolo:2018eco,Reina:2015gxa,Deruelle:2007pt,Padilla:2012ze,Senovilla,Macias:2002sr}, we analyze the junction conditions for $f(\mathcal{Q})$-gravity within the framework of distribution--valued tensors \cite{Dray1,Dray2}. Borrowing arguments and notations from \cite{Poisson}, we deduce the necessary and sufficient conditions for two given solutions can be smoothly joined in the distributional sense, so having a smooth transition at the level of field equations as well. Our findings are summarized into four Propositions concerning the cases of null and non--null hypersurface separately, under two distinct assumptions: the request for regularity of the Riemann tensor across the separation hypersurface or no requirement concerning the regularity of the Riemann tensor. If the separation hypersurface is timelike or spacelike, the necessary and sufficient conditions coincide. Instead, for null hypersurfaces, they differ slightly. As expected, compared to GR, new conditions arise that explicitly involve the nonmetricity tensor.

Section \ref{Section_II} briefly reviews some generalities about $f(\mathcal{Q})$-gravity. Section \ref{Section_III} discusses junction conditions. Section \ref{Section_IV} is devoted to conclusions. Throughout the paper natural units ($c=8\pi G=1$) and metric signature ($-,+,+,+$) are used.
\section{\texorpdfstring{$f(\mathcal{Q})$}{}-gravity}\label{Section_II}
Let $\cal M$ be a spacetime manifold endowed with a metric tensor $g_{ij}$ and an affine connection $\Gamma_{ij}{}^{k}$ which is assumed to be torsion free ($S_{ij}{}^{k}:=\Gamma_{ij}{}^{k}-\Gamma_{ji}{}^{k}=0$). After introducing the nonmetricity tensor
\begin{equation}\label{eq:def_nonmetricity}
Q_{kij} = \nabla_{k}g_{ij}
\end{equation}
where $\nabla_k$ is the covariant derivative associated with the given affine connection $\Gamma_{ij}{}^{k}$, we can decompose the affine connection itself as
\begin{equation}\label{eq:def_connection}
\Gamma_{ij}{}^{k} = \tilde{\Gamma}_{ij}{}^{k} + N_{ij}{}^{k} 
\end{equation}
In eq. \eqref{eq:def_connection} $\tilde{\Gamma}_{ij}{}^{k}$ denotes the Levi-Civita connection induced by the metric tensor $g_{ij}$
\begin{equation}\label{eq:def_levicivita}
\tilde{\Gamma}_{ij}{}^{k} = \frac{1}{2} g^{kh} \left( \de{g_{jh}}/de{x^i} + \de{g_{ih}}/de{x^j} - \de{g_{ij}}/de{x^h} \right)
\end{equation}
whereas $N_{ij}{}^{k}$ indicates the disformation tensor
\begin{equation}\label{eq:def_disformation}
N_{ij}{}^{k} = \frac{1}{2} \left( Q^{k}{}_{ij} - Q_{i}{}^{k}{}_{j} - Q_{j}{}^{k}{}_{i} \right)
\end{equation}
Making use of the two distinct traces of the nonmetricity tensor
\begin{equation}\label{traces_Q}
    q_{h} = Q_{hi}{}^{i} \quad \mbox{and} \quad Q_{h}=Q_{ih}{}^{i}.
\end{equation}
the nonmetricity scalar can be defined as
\be\label{scalar_Q}
\mathcal{Q} = \frac{1}{4}Q_{hij}Q^{hij} - \frac{1}{2}Q_{hij}Q^{ijh} - \frac{1}{4} q_{h}q^{h} + \frac{1}{2}q_{h}Q^{h}
\ee
The latter can also be expressed as
\be\label{scalar_qbis}
\mathcal{Q}=-Q_{hij}P^{hij}
\ee
after introducing the tensor
\be\label{3.1.1bis}
P^{h}{}_{ij} = - \frac{1}{4} Q^{h}{}_{ij} + \frac{1}{4} Q_{ij}{}^{h} + \frac{1}{4} Q_{ji}{}^{h} + \frac{1}{4} q^{h} g_{ij} - \frac{1}{4} Q^{h} g_{ij} - \frac{1}{8} \delta^{h}_{i} q_{j} - \frac{1}{8}\delta^{h}_{j} q_{i}
\ee
$f(Q)$-gravity is a metric--affine theory of gravitation whose gravitational Lagrangian is a function $f(Q)$ of the nonmetricity scalar. More in detail, the action functional is given by
\be
\mathcal{A}=\int \left[ -\frac{1}{2}\sqrt{-g} f(\mathcal{Q}) + \lambda_{h}{}^{kij}R^{h}{}_{kij} + \lambda_{h}{}^{ij}S_{ij}{}^{h} + \mathcal{L}_m\right]d^4x 
\ee
where $f(\mathcal{Q})$ is a given function of the nonmetricity scalar,
$\mathcal{L}_{m}$ indicates the matter Lagrangian density, $R^{h}{}_{kij}$ and $S_{ij}{}^{k}$ are the curvature and the torsion of the dynamic connection, and $\lambda_{a}{}^{bij}$ and $\lambda_{a}{}^{ij}$ are Lagrange multipliers. By varying with respect to the Lagrange multipliers, we obtain the constraints
\begin{equation}\label{curvature_equations}
R^{h}{}_{kij} = 0 \qquad \mbox{and} \qquad S_{ij}{}^{h} = 0
\end{equation}
which imply that the dynamic connection is flat and torsionless. Variations with respect to the metric and the connection yield field equations of the form
\begin{equation} \label{eq:metric_equation}
\begin{split}
\frac{2}{\sqrt{-g}}\nabla_{h} \left( \sqrt{-g} ft P^{h}{}_{ij} \right) + \frac{1}{2}g_{ij}f(\mathcal{Q}) 
+ f' \left( P_{iab}Q_{j}{}^{ab} - 2 Q^{ab}{}_{i}P_{abj} \right) = T_{ij}
\end{split}
\end{equation}
and
\begin{equation}\label{eq:connection_equation}
 \nabla_{p} \lambda_{h}{}^{jip} +  \lambda_{h}{}^{ij} - \sqrt{-g} f' P^{ij}{}_{h} = \Phi^{ij}{}_{h},
\end{equation}
where the quantities 
\begin{equation}
    T_{ij} :=  -\frac{2}{\sqrt{-g}}\frac{\delta \mathcal{L}_{m}}{\delta g^{ij}} \quad \text{and} \quad
    \Phi^{ij}{}_{h} := - \frac{1}{2}\frac{\delta \mathcal{L}_{m}}{\delta {\Gamma_{ij}{}^{h}}}
\end{equation}
are the energy-momentum and the hypermomentum tensors, respectively. In the case the hypermomentum is zero, the usual conservation laws
\be\label{conservation_laws} 
\tilde{\nabla}_j\/T^{ij}
\ee
hold, where $\tilde{\nabla}$ denotes the Levi--Civita covariant derivative. Moreover, the constraints \eqref{curvature_equations} ensure the existence of local coordinates in which the connection coefficients are zero. The choice of the dynamic connection is then a pure gauge. Once this choice has been made, eqs. \eqref{eq:connection_equation} are intended for the determination of the Lagrangian multipliers only. On the other hand, Lagrangian multipliers do not enter the field equations \eqref{eq:metric_equation}, and then their determination can be omitted together with eqs. \eqref{eq:connection_equation}. To conclude, it is worth noticing that after separating Levi-Civita contributions from nonmetricity ones, the field equations \eqref{eq:metric_equation} can be recast in the Einstein-like form 
\begin{equation}\label{einstein_equations}
\tilde{G}_{ij} = \frac{1}{f'}T_{ij} - \frac{1}{2}g_{ij}\left(\frac{f}{f'} - \mathcal{Q}\right) - 2\frac{f''}{f'}P^{h}{}_{ij}\partial_{h}\mathcal{Q}
\end{equation} 
where $\tilde{G}_{ij}$ is the Einstein tensor generated by the Levi-Civita connection of the metric $g_{ij}$. In general, hereafter, we will denote by the tilde symbol all quantities associated with the Levi-Civita connection.
\section{Junction conditions}\label{Section_III}
We discuss the problem of joining together, on a given hypersurface $\Sigma$, two different solutions $\left(g_{ij}^+,\Gamma_{ij}^{+\;\;h}\right)$ and $\left(g_{ij}^-,\Gamma_{ij}^{-\;\;h}\right)$ of the field equations \eqref{curvature_equations} and \eqref{einstein_equations}. We imagine that $\Sigma$ separates two distinct regions ${\cal M}^+$ and ${\cal M}^-$ of spacetime, where the two solutions are respectively defined. First, we assume $\Sigma$ to be either timelike or spacelike hypersurface. We will address the case of null hypersurface after.

Borrowing ideas and notations from \cite{Poisson}, we deal with the problem of junction conditions in the framework of distribution--valued tensors \cite{L1,L2,Choquet,Dray1,Dray2}. To this end, we preliminarily introduce a coordinate system $x^i$, locally overlapping both ${\cal M}^+$ and ${\cal M}^-$ in a neighborhood of $\Sigma$. Also, we refer the hypersurface $\Sigma$ to local coordinates $y^A$ ($A=1,\ldots,3$). Moreover, we denote by 
\begin{equation}
\left[W\right] := W\left({\cal M}^+\right)_{|\Sigma} - W\left({\cal M}^-\right)_{|\Sigma}
\end{equation}
the jump across $\Sigma$ of any geometric quantity $W$ defined on both sides of the hypersurface.

After that, we need a function over spacetime whose sign allows us to distinguish ${\cal M}^+$ from ${\cal M}^-$. The latter can be defined by considering the arc's length $s$ between any point $p\in {\cal M}$ and $\Sigma$, measured along the geodesic normal to $\Sigma$ and passing through $p$. For such geometrical construction, one of the two metrics needs to be chosen, for instance, $g_{ij}^+$. Without loss of generality, we can suppose to have $s<0$ in ${\cal M}^-$, $s>0$ in ${\cal M}^+$ and $s=0$ at $\Sigma$. Denoting by $n^i$ the unit normal (with respect to the chosen metric) outgoing from $\Sigma$, the following identities hold
\begin{equation}\label{3.0}
dx^i = n^i\/ds, \qquad n_i = \epsilon\frac{\partial s}{\partial x^i} \qquad{\rm and}\qquad n^i\/n_i =\epsilon 
\end{equation}
where $\epsilon =1$ if $\Sigma$ is timelike, $\epsilon =-1$ if $\Sigma$ is spacelike. 
 
Now, introducing the Heaviside distribution $\Theta(s)$ (with $\Theta(0):=1$) and taking into account the two solutions $\left(g_{ij}^+,\Gamma_{ij}^{+\;\;h}\right)$ and $\left(g_{ij}^-,\Gamma_{ij}^{-\;\;h}\right)$ as above, we define the distribution--valued quantities
\begin{subequations}\label{3.1}
\begin{equation}\label{3.1a}
g_{ij} = \Theta(s)\/g_{ij}^+ + \left(1-\Theta(s)\right)\/g_{ij}^-
\end{equation}
\begin{equation}\label{3.1b}
\Gamma_{ij}^{\;\;\;h} = \Theta(s)\Gamma_{ij}^{+\;\;h} + \left(1-\Theta(s)\right)\Gamma_{ij}^{-\;\;h} 
\end{equation}
\end{subequations}
We want to study the conditions under which the quantities \eqref{3.1} represent a solution of the field equations \eqref{curvature_equations} and \eqref{einstein_equations} on the entire spacetime in a distributional sense. At the same time, denoting by $\tilde{\Gamma}_{ij}^{+\;\;h}$ and $\tilde{\Gamma}_{ij}^{-\;\;h}$ the Levi--Civita connections defined in ${\cal M}^+$ and ${\cal M}^-$ respectively, we also require that the quantity
\begin{equation}\label{3.1_connection}
\tilde{\Gamma}_{ij}^{\;\;\;h} = \Theta(s)\tilde{\Gamma}_{ij}^{+\;\;h} + \left(1-\Theta(s)\right)\tilde{\Gamma}_{ij}^{-\;\;h} 
\end{equation}
be the Levi--Civita connection associated with the metric \eqref{3.1a}, and  
\be\label{3.1.2}
\tilde{R}^h_{\;\,kij} = \Theta(s)\tilde{R}^{+h}_{\;\;\;\;kij} + \left(1-\Theta(s)\right)\tilde{R}^{-h}_{\;\;\;\;kij} 
\ee
be its Riemann tensor, i.e., the curvature tensor induced by the connection \eqref{3.1_connection}, still in a distributional sense. Hereafter, we will use the term {\it Riemann tensor} exclusively to indicate the curvature tensor of the Levi--Civita connection. In eq. \eqref{3.1.2}, $\tilde{R}^{+h}_{\;\;\;\;kij}$ and $\tilde{R}^{-h}_{\;\;\;\;kij}$ are then the Riemann tensors in ${\cal M}^+$ and ${\cal M}^-$ respectively. Condition \eqref{3.1.2} is noting else the requirement of regularity across the separation hypersurface $\Sigma$ for the Riemann tensor $\tilde{R}^h_{\;\,kij}$. For instance, this is consistent with the interpretation of $f(\mathcal Q)$-gravity, according to which nonmetricity generates modified Einstein equations via a suitable variational principle, but the gravitational field continues to be represented by the Levi-Civita connection $\tilde{\Gamma}_{ij}^{\;\;\;h}$. By this, we mean that the free--fall motions are the geodesics of the Levi--Civita connection, so gravitation remains connected to the Riemann tensor $\tilde{R}^h_{\;\,kij}$, hence the requirement for its regularity. The above interpretation reflects the ideas of Refs. [10,12-14], but it should be pointed out that in literature these aspects are still matter of debate and other points of view exist. 

The first condition that must be satisfied is the consistency of \eqref{3.1} with the kinematic equation \eqref{eq:def_nonmetricity}. In this regard, inserting eqs. \eqref{eq:def_connection} and \eqref{3.1b} into eq. \eqref{eq:def_nonmetricity} and separating all the contributions due to the Levi--Civita connection from those due to nonmetricity, we get the relation
\begin{equation}\label{3.1bis}
\nabla_h\/g_{ij} = \Theta(s)Q^+_{ijh} + \left(1-\Theta(s)\right)Q^-_{ijh} + \epsilon\/n_h\left[g_{ij}\right]\delta(s)
\end{equation}
where the identities $\de{s}/de{x^i}= \epsilon\/n_i$, $\frac{d\Theta(s)}{ds}=\delta(s)$, $\Theta^2(s)=\Theta(s)$ and $\Theta(s)\left(1-\Theta(s)\right)=0$ have been used. The required consistency is achieved if the term involving the Dirac $\delta$-function vanishes. This implies the condition
\begin{equation}\label{3.1tris}
\left[g_{ij}\right] =0
\end{equation} 
amounting to the continuity of the two metric tensors $g^+_{ij}$ and $g^-_{ij}$ across the hypersurface $\Sigma$. Condition \eqref{3.1tris} also guarantees consistency of eq. \eqref{3.1a} with the definition of the Christoffel coefficients \eqref{eq:def_levicivita}. As a matter of fact, evaluating the Christoffel coefficients of the metric \eqref{3.1a}, one gets
\begin{equation}\label{3.2}
\tilde{\Gamma}_{ijh} = \Theta(s)\tilde{\Gamma}^+_{ijh} + \left(1-\Theta(s)\right)\tilde{\Gamma}^-_{ijh} 
\end{equation}
where the identity $\de{g_{ij}}/de{x^k} = \Theta(s)\de{g_{ij}^+}/de{x^k} + \left(1-\Theta(s)\right)\de{g_{ij}^-}/de{x^k} + \epsilon\/\delta(s)\left[g_{ij}\right]\/n_k$ has been systematically used together with eq. \eqref{3.1tris}.

Further junction conditions can be derived by requiring smooth transition across $\Sigma$ at the level of the dynamic equations \eqref{curvature_equations} and \eqref{einstein_equations}. To see this point in detail, we start by implementing eqs. \eqref{curvature_equations}, evaluated for the affine connection \eqref{3.1b}. Making use of the identity $\de{\Gamma_{ij}^{\;\;\;h}}/de{x^k} = \Theta(s)\de{\Gamma_{ij}^{+\;\;h}}/de{x^k} + \left(1-\Theta(s)\right)\de{\Gamma_{ij}^{-\;\;h}}/de{x^k} + \epsilon\delta(s)\/\left[\Gamma_{ij}^{\;\;\;h}\right]n_k$, we have 
\begin{equation}\label{3.3}
R^{p}_{\;\;qij} = \Theta(s)R^{+p}_{\;\;\;\;qij} + \left(1-\Theta(s)\right)R^{-p}_{\;\;\;\;qij} + \delta(s)\/A^{p}_{\;\;qij} =  \delta(s)\/A^{p}_{\;\;qij} = 0
\end{equation}
where
\begin{equation}\label{3.4}
A^{p}_{\;\;qij} := \epsilon\left(\left[\Gamma_{jq}^{\;\;\;p}\right]\/n_i - \left[\Gamma_{iq}^{\;\;\;p}\right]\/n_j\right)
\end{equation}
The field equations for the curvature of the dynamic connection $\Gamma_{ij}^{\;\;\;h}$ are then satisfied if and only if the tensor $A^{p}_{\;\;qij}$ vanishes. In order to discuss this condition, it is convenient to separate the Levi--Civita contributions from the nonmetricity ones. Given eq. \eqref{eq:def_connection}, we easily obtain
\begin{equation}\label{3.5}
A^{p}_{\;\;qij} = \tilde{A}^{p}_{\;\;qij} + \bar{A}^{p}_{\;\;qij}
\end{equation}
where
\begin{equation}\label{3.6}
\tilde{A}^{p}_{\;\;qij} = \epsilon\left(\left[\tilde{\Gamma}_{jq}^{\;\;\;p}\right]\/n_i - \left[\tilde{\Gamma}_{iq}^{\;\;\;p}\right]\/n_j\right)
\end{equation}
is the quantity due to the Levi--Civita connection, and
\begin{equation}\label{3.7}
\bar{A}^{p}_{\;\;qij} = \epsilon\left(\left[N_{jq}^{\;\;\;p}\right]\/n_i - \left[N_{iq}^{\;\;\;p}\right]\/n_j\right)
\end{equation}
is the quantity due to nonmetricity. If in addition to this we require that the Riemann tensor $\tilde{R}^{p}_{\;\;qij}$ associated with the metric \eqref{3.1a} is regular on the separation hypersurface $\Sigma$ as well, both quantities $\tilde{A}^{p}_{\;\;qij}$ and $\bar{A}^{p}_{\;\;qij}$ must vanish separately.

As for the quantity $\tilde{A}^{p}_{\;\;qij}$, its analysis is closely connected with the requirement of smooth transition across $\Sigma$ for the Einstein-like equations \eqref{einstein_equations}. The discussion on this point is very similar to that given in \cite{Poisson} for Einstein equations in General Relativity. Therefore, here we recall only the main steps and outcomes for brevity, referring the reader to \cite{Poisson} for more details. 

The jump of the partial derivatives of the metric can be expressed using a $2$-rank symmetric tensor $k_{ij}$ on $\Sigma$, in such a way that one has
\begin{equation}\label{3.8}
\left[\frac{\partial g_{ij}}{\partial x^h}\right] = k_{ij}\/n_h
\end{equation}
Making use of \eqref{3.8}, we obtain the expressions
\begin{equation}\label{3.10}
\left[\tilde{\Gamma}_{ij}^{\;\;\;h}\right] = \frac{1}{2}\left(k^h_{\;\;j}\/n_i + k^h_{\;\;i}\/n_j - k_{ij}\/n^h\right)
\end{equation}
by which we get the explicit representation
\begin{equation}\label{3.11}
\tilde{A}^{p}_{\;\;qij} = \frac{\epsilon}{2}\left(k^p_{\;\;j}\/n_q\/n_i - k^p_{\;\;i}\/n_q\/n_j - k_{qj}\/n^p\/n_i + k_{qi}\/n^p\/n_j\right)
\end{equation}
The expression \eqref{3.11} enables us to calculate the Einstein tensor, which is associated with the (distribution--valued) metric tensor \eqref{3.1a}
\begin{equation}\label{3.11bis}
\tilde{G}_{ij} = \Theta(s)\tilde{G}^+_{ij} + \left(1-\Theta(s)\right)\tilde{G}^-_{ij} + \delta(s)\/\tilde{S}_{ij}
\end{equation}
where $\tilde{G}^+_{ij}$ and $\tilde{G}^-_{ij}$ are the Einstein tensors induced by the metric tensors $g^+_{ij}$ and $g^-_{ij}$ respectively, and
\begin{equation}\label{3.14}
\tilde{S}_{ij} := \frac{\epsilon}{2}\left(k^p_{\;\;j}\/n_i\/n_p - k\/n_i\/n_j - k_{ij}\epsilon + k_{ip}\/n^p\/n_j\right) - \frac{\epsilon}{2}\left(k_{st}\/n^s\/n^t - \epsilon\/k\right)\/g_{ij}
\end{equation} 
By construction, the tensor \eqref{3.14} is symmetric and tangent to the hypersurface $\Sigma$. Denoting by $E^i_A :=\de{x^i}/de{y^A}$ the components of the vectors tangent to the coordinate lines of the surface $\Sigma$, $\tilde{S}_{ij}$ possesses an equivalent representation of the form 
\begin{equation}\label{3.15}
\tilde{S}_{AB} := \tilde{S}_{qj}\/E^q_A\/E^j_B = -\frac{1}{2}k_{qj}\/E^q_A\/E^j_B + \frac{1}{2}k_{pq}h^{pq}\/h_{AB}
\end{equation}
where $h^{pq}:= g^{pq} -\epsilon\/n^p\/n^q$ and $h_{AB}:=g_{ij}\/E^i_A\/E^j_B$ are the projection operator and the induced metric on $\Sigma$ respectively. In connection with this, after introducing the extrinsic curvature
\begin{equation}\label{3.16}
K_{AB} := \left(\tilde{\nabla}_i\/n_j\right)\/E^j_A\/E^i_B 
\end{equation}
we have the identity
\begin{equation}\label{3.17}
\tilde{S}_{AB} := - \epsilon\left(\left[K_{AB}\right] - \left[K\right]\/h_{AB}\right) 
\end{equation}
where $\left[K\right]:= \left[K_{AB}\right]\/h^{AB}$. The conclusion follows that the vanishing of the term $\tilde{S}_{ij}$ amounts to the vanishing of the jump of the extrinsic curvature. A smooth transition across $\Sigma$ of the Einstein-like equations \eqref{einstein_equations} implies then that 
\begin{equation}\label{3.18}
\left[K_{AB}\right] = \frac{\epsilon}{2}k_{ij}E^i_AE^j_B = 0
\end{equation}
In addition to removing the $\delta$-function term from the left-hand side of the Einstein-like equations, condition \eqref{3.18} ensures that $\tilde{A}^{p}_{\;\;qij}=0$ as well. This can be verified by evaluating the tensor \eqref{3.11} on the tensor products of the elements of the basis $\{n^i,E^i_A\}$. In such a circumstance, the Riemann tensor generated by the metric \eqref{3.1a} is nonsingular at $\Sigma$. Moreover, if condition \eqref{3.18} holds, the vanishing of tensor \eqref{3.5} reduces to the vanishing of \eqref{3.7}, thus
\begin{equation}\label{3.19}
\left(\left[N_{jq}^{\;\;\;p}\right]\/n_i - \left[N_{iq}^{\;\;\;p}\right]\/n_j\right) = 0
\end{equation}
Furthermore, denoting by
\begin{equation}\label{3.20}
\bar{T}_{ij} := \frac{1}{f'}T_{ij} - \frac{1}{2}g_{ij}\left(\frac{f}{f'} - \mathcal{Q}\right) - 2\frac{f''}{f'}P^{h}{}_{ij}\partial_{h}\mathcal{Q}
\end{equation}
the effective energy-momentum tensor on the right-hand side of \eqref{einstein_equations}, we have the identity
\begin{equation}\label{3.21}
\Theta(s)\tilde{G}^+_{ij} + \left(1-\Theta(s)\right)\tilde{G}^-_{ij} = \Theta(s)\/\bar{T}^+_{ij} + \left(1-\Theta(s)\right)\/\bar{T}^-_{ij}
\end{equation}
where the left hand side of \eqref{3.21} is just the distribution--valued Einstein tensor induced by the metric \eqref{3.1a} when condition \eqref{3.1tris} and \eqref{3.18} hold. Consistency with the distributional approach would require that the right-hand side of eq. \eqref{3.21}  coincide with the analogous term calculated starting from the distributional solution and in particular from $P^h{}_{ij}=\Theta(s)P^{+h}{}_{ij} + \left(1-\Theta(s)\right)P^{-h}{}_{ij}$ and $\mathcal{Q}=\Theta(s)\mathcal{Q}^+ + \left(1-\Theta(s)\right)\mathcal{Q}^-$. In order for this request to be satisfied, the following conditions have to hold
\begin{equation}\label{3.22}
\left(f''\right)^+_{|\Sigma}=0 \quad\cup\quad {P^{+h}{}_{ij}n_h}_{|\Sigma}=0 \quad\cup\quad \left[\mathcal{Q}\right] = 0
\end{equation}
This is because the $\delta$-function term which arises on the right-hand side of the field equations \eqref{einstein_equations}, and which has to be zero, is of the form
\be\label{3.22bis}
\delta(s)\epsilon\left(-2\frac{f''}{f'}\right)^+P^{+h}{}_{ij}n_h\left[\mathcal{Q}\right]
\ee
It is worth noticing that the term \eqref{3.22bis} follows from defining the Heaviside function at zero as $\Theta(0)=1$. By defining $\Theta(0)=0$, we would get a similar expression but involving quantities evaluated on ${\cal M}^-$. On the other hand, when discussing junction conditions, other authors are used to adopting a not-defined Heaviside function at zero. If we followed this convention, the singular term arising from the effective energy-momentum tensor \eqref{3.20} would be of the form 
\be\label{3.22tris}
-2\epsilon\left\{ \Theta(s)\/\left(\frac{f''}{f'}\right)^{+}P^{+h}{}_{ij} + \left(1-\Theta(s)\right)\left(\frac{f''}{f'}\right)^{-}P^{-h}{}_{ij}\right\}n_h\left[\mathcal{Q}\right]\delta(s)
\ee
and it would not be well defined as a distribution-valued tensor due to the presence of the products $\Theta\delta$. Therefore, it would be ambiguous to set the term \eqref{3.22tris} equal to zero when the latter is not well defined. This ambiguity is easily overcome by defining the Heaviside function at zero. Furthermore, defining the Heaviside function at zero allows us to give full meaning to the term \eqref{3.22bis}, which is interpretable as the energy-momentum tensor of the thin shell separating ${\cal M}^+$ from ${\cal M}^-$ when conditions \eqref{3.22} are not met.

Summarizing all the obtained results, we may state the following
\\
{\Proposition Given a timelike or spacelike hypersurface $\Sigma$, requirements \eqref{3.1tris}, \eqref{3.18}, \eqref{3.19} and \eqref{3.22} are necessary and sufficient conditions for the quantities \eqref{3.1} to be well-posed solutions of the field equations \eqref{curvature_equations} and \eqref{einstein_equations} in the distributional sense, and for the Riemann tensor induced by metric \eqref{3.1a} to be nonsingular on the separation hypersurface $\Sigma$.}
\\
\\
As a further remark, it is worth noticing that the absence of the $\delta$-function term in the Einstein equations \eqref{3.11bis} allows us to recast them in the equivalent form
\begin{equation}\label{3.23}
\Theta(s)\tilde{R}^+_{ij} + \left(1-\Theta(s)\right)\tilde{R}^-_{ij} = \Theta(s)\left(\bar{T}^+_{ij} -\frac{1}{2}\bar{T}^+\/g^+_{ij}\right) + \left(1-\Theta(s)\right)\left(\bar{T}^-_{ij} -\frac{1}{2}\bar{T}^-\/g^-_{ij}\right)
\end{equation}
$\tilde{R}_{ij}$ denoting the Ricci tensor and $\bar{T}:=\bar{T}_{ij}g^{ij}$ the usual trace. Additionally, by saturating \eqref{3.19} with $h^i_s\/n^j$ (where $h^i_s:= \delta^i_s - \epsilon\/n^i\/n_s$) we get necessary conditions
\be\label{necessary_conditions}
\left[N_{sk}^{\;\;\;h}\right] = \left[N_{ik}^{\;\;\;h}\right]n^i\/n_s\/\epsilon \quad \Rightarrow \quad \left[N_{sk}^{\;\;\;h}\right]h^s_i = \left[N_{ks}^{\;\;\;h}\right]h^s_i = 0
\ee
to be satisfied by the jump of the disformation tensor. 

We carried out the previous discussion under the requirement that also the Riemann tensor $\tilde{R}^h_{\;\,kij}$ be regular on the separation hypersurface. This condition is very strong and leads to findings very similar to those of GR. Actually, we can study junction conditions even without adopting such an assumption. Indeed, making use of the previous calculations, it is easily seen that the vanishing of the $\delta$-term arising in the Einstein--like equations \eqref{einstein_equations} yields the set of equations
\begin{subequations}\label{4.1}
\be\label{4.1a}
\left[K_{AB}\right] - \left[K\right]\/h_{AB}  = 2\left(\frac{f''}{f'}\right)^+P^{+h}{}_{ij}n_hE^i_AE^j_B\left[\mathcal{Q}\right]
\ee
\be\label{4.1b}
\left(\frac{f''}{f'}\right)^+P^{+h}{}_{ij}n_hn^in^j\left[\mathcal{Q}\right] =0, \quad
\left(\frac{f''}{f'}\right)^+P^{+h}{}_{ij}n_hE^i_An^j\left[\mathcal{Q}\right] =0
\ee
\end{subequations}
where we used the fact that the tensor $\tilde{S}_{ij}$ is tangent to the hypersurface $\Sigma$ and where now the two sides of eq. \eqref{4.1a} are not forced to be separately zero. In particular, eq. \eqref{4.1a} can be recast in the equivalent form
\be\label{4.2}
\left[K_{AB}\right] = 2\left(\frac{f''}{f'}\right)^+P^{+h}{}_{ij}n_h\left[\mathcal{Q}\right]\left(E^i_AE^j_B - \frac{1}{2}h^{ij}h_{AB} \right)
\ee
which expresses the jump of the extrinsic curvature in terms of contributions due to nonmetricity and the function $f(\mathcal Q)$. After that, the field equations \eqref{curvature_equations} impose the vanishing of the tensor \eqref{3.4} which generates the equations
\be\label{4.3}
\left[\tilde{\Gamma}_{jq}^{\;\;\;p}\right]\/n_i - \left[\tilde{\Gamma}_{iq}^{\;\;\;p}\right]\/n_j = -\left[N_{jq}^{\;\;\;p}\right]\/n_i + \left[N_{iq}^{\;\;\;p}\right]\/n_j
\ee
relating the jumps of the Levi--Civita connection to the jumps of the disformation tensor. Once again, we can summarize our findings by the following
\\
{\Proposition Given a timelike or spacelike hypersurface $\Sigma$, requirements \eqref{4.1b}, \eqref{4.2} and \eqref{4.3} are necessary and sufficient conditions for the quantities \eqref{3.1} to be well--posed solutions of the field equations \eqref{curvature_equations} and \eqref{einstein_equations} in the distributional sense.}
\\
\\
Thus, when the regularity requirement for the Riemann tensor is removed, we obtain junction conditions where Levi--Civita contributions are coupled to nonmetricity ones. In particular, eq. \eqref{4.2} implies that the jump of the extrinsic curvature is related to the jump of the nonmetricity scalar together with the values of the second derivative $f''$ and the tensor $P^{h}{}_{ij}$ at $\Sigma$. In addition, eq. \eqref{4.3} shows that the jumps of the Levi--Civita connection are related to the jumps of the disformation tensor. The above are relevant differences compared to the case in which the regularity of the Riemann tensor is imposed.

We discuss now the case in which $\Sigma$ is a null hypersurface, described by an equation of the form $\Phi(x^i)=0$. Borrowing again from \cite{Poisson}, in such a circumstance the normal vector $n_i$ can be expressed as
\be\label{3.24}
n_i = \alpha^{-1}\partial_i \Phi
\ee
where $\alpha$ is a suitable non--zero function on $\Sigma$. Without loss of generality, we can suppose that ${\cal M}^+$ and ${\cal M}^-$ correspond to the domains where $\Phi$ is positive and negative, respectively. We refer $\Sigma$ to local coordinates $(\lambda,y^A)$, $A=1,2$, in such a way that $n^i=\de{x^i}/de{\lambda}$ and $n_iE^i_A=0$, with $E^i_A:=\de{x^i}/de{y^A}$. The matching on $\Sigma$ of two solutions of the field equations is given by 
\begin{subequations}\label{3.1bb_null}
\begin{equation}\label{3.1bba_null}
g_{ij} = \Theta(\Phi)\/g_{ij}^+ + \left(1-\Theta(\Phi)\right)\/g_{ij}^-, 
\end{equation}
\begin{equation}\label{3.1bbb_null}
\Gamma_{ij}^{\;\;\;h} = \Theta(\Phi)\Gamma_{ij}^{+\;\;h} + \left(1-\Theta(\Phi)\right)\Gamma_{ij}^{-\;\;h}. 
\end{equation}
\end{subequations}
By repeating similar reasoning as in previous timelike or spacelike cases, the kinematic equation \eqref{eq:def_nonmetricity} yields again the continuity condition for the metric tensor, $[g_{ij}]=0$. To proceed further, we need to introduce a transverse vector field $N^i$ on $\Sigma$, satisfying the requirements $N^in_i=-1$, $N^iN_i=0$, $N_iE^i_A=0$ and complementing $\{n^i,E^i_A\}$ to a basis of $T_\Sigma\/M$. After that, we also introduce the transverse metric
\begin{equation}\label{transversemetric}
h_{ij} = g_{ij} + n_iN_j + n_jN_i. 
\end{equation} 
The curvature tensor of the connection \eqref{3.1bbb_null} is of the form
\begin{equation}\label{3.3_null}
R^{p}_{\;\;qij} = \delta(\Phi)\left(\tilde{A}^{p}_{\;\;qij} + \bar{A}^{p}_{\;\;qij}\right)
\end{equation}
with
\be\label{3.25}
\tilde{A}^{p}_{\;\;qij} = \alpha\left(\left[\tilde{\Gamma}_{jq}^{\;\;\;p}\right]\/n_i - \left[\tilde{\Gamma}_{iq}^{\;\;\;p}\right]\/n_j\right) \quad {\rm and} \quad \bar{A}^{p}_{\;\;qij} = \alpha\left(\left[N_{jq}^{\;\;\;p}\right]\/n_i - \left[N_{iq}^{\;\;\;p}\right]\/n_j\right)
\ee
Once again, we analyze the two contributions \eqref{3.25} separately. As for the term $\tilde{A}^{p}_{\;\;qij}$, the continuity across $\Sigma$ implies that the derivatives of the metric tensor may have discontinuities only along the transverse direction, compared with the other directions given by the vectors $\{n^i,E^i_A\}$. So we can infer the existence of a tensor field $\gamma_{ij}$ on $\Sigma$, such that
\begin{equation}\label{discontinuitytensor}
\gamma_{ij} = - \left[\de{g_{ij}}/de{x^s}\right]N^s \qquad\Longleftrightarrow\qquad \left[\de{g_{ij}}/de{x^s}\right] = \gamma_{ij}n_s. 
\end{equation}
In view of eq. \eqref{discontinuitytensor}, the jump of the Christoffel symbols can be expressed as
\begin{equation}\label{3.10bbnull}
\left[\tilde{\Gamma}_{ij}^{\;\;\;h}\right] = \frac{1}{2}\left(\gamma^h_{\;\;j}\/n_i + \gamma^h_{\;\;i}\/n_j - \gamma_{ij}\/n^h\right). 
\end{equation}
Replacing \eqref{3.10bbnull} in the first of \eqref{3.25}, we get the representation
\be\label{3.26}
\tilde{A}^{p}_{\;\;qij} =\frac{\alpha}{2}\left(\gamma^p_{\;\;j}n_qn_i - \gamma^p_{\;\;i}n_qn_j - \gamma_{jq}n^pn_i + \gamma_{iq}n^pn_j\right)
\ee
and then identity
\be\label{3.27}
\tilde{A}_{ij} - \frac{1}{2}\tilde{A}g_{ij} = \frac{\alpha}{2}\left(\gamma_{ih}n^hn_j + \gamma_{jh}n^hn_i - \gamma^h_hn_in_j - \gamma_{hk}n^hn^kg_{ij}\right)
\ee
Tensor \eqref{3.27} is the term responsible for the presence of the $\delta$-function singularity in the Einstein tensor generated by the distribution-valued metric \eqref{3.1bba_null}. Therefore, a smooth transition across the hypersurface $\Sigma$ at the level of the Einstein-like equations needs the following identity to hold on to the separation hypersurface $\Sigma$
\be\label{3.27bis}
\tilde{A}_{ij} - \frac{1}{2}\tilde{A}g_{ij} = \alpha\left(-2\frac{f''}{f'}\right)^+P^{+h}{}_{ij}n_h\left[\mathcal{Q}\right]
\ee
If we require that the Riemann tensor, induced by \eqref{3.1a}, is smooth across $\Sigma$ and that the same holds for the corresponding Einstein tensor, both sides of eq. \eqref{3.27bis} have to vanish separately. In particular,  tensor \eqref{3.27} can be decomposed as 
\be\label{3.28}
\tilde{A}_{ij} - \frac{1}{2}\tilde{A}g_{ij} = \frac{\alpha}{2}\left\{-\gamma_{hk}n^hn^kh_{ij} + \gamma_{sh}n^h\left(h^s_jn_i + h^s_in_j\right) - \gamma_{hk}h^{hk}n_in_j\right\}
\ee
The conclusion follows that the vanishing of \eqref{3.27} amounts to the vanishing of the projections
\be\label{3.29}
\gamma_{hk}n^hn^k=0, \quad \gamma_{sh}n^hh^s_j=0, \quad \gamma_{hk}h^{hk}=0
\ee
The quantities in eq. \eqref{3.29} are connected with the jump of the transverse curvature, defined as the projection on $\Sigma$ of the covariant derivative $\tilde{\nabla}_iN_j$. Indeed, given $C_{\lambda\lambda}:=\left(\tilde{\nabla}_iN_j\right)n^in^j$, $C_{ij}:=\left(\tilde{\nabla}_sN_t\right)h^s_ih^t_j$, $C_{i\lambda}:=\left(\tilde{\nabla}_sN_t\right)h^s_in^t$ and $C_{\lambda\/i}:=\left(\tilde{\nabla}_sN_t\right)h^t_in^s$ we have
\be\label{3.30}
\left[C_{\lambda\lambda}\right]=-\frac{1}{2}\gamma_{hk}n^hn^k, \quad \left[C_{ij}\right]=\left[C_{ji}\right]=-\frac{1}{2}\gamma_{hk}h^{h}_ih^k_j, \quad \left[C_{i\lambda}\right]=\left[C_{\lambda\/i}\right]=-\frac{1}{2}\gamma_{sh}n^hh^s_i
\ee
Clearly, the vanishing of the quantities \eqref{3.30}, namely 
\be\label{3.31}
\left[C_{\lambda\lambda}\right]=0, \quad \left[C_{ij}\right]=\left[C_{ji}\right]=0, \quad \left[C_{i\lambda}\right]=\left[C_{\lambda\/i}\right]=0
\ee
ensures that conditions \eqref{3.29} are satisfied. Moreover, whenever conditions \eqref{3.31} are satisfied, the tensor \eqref{3.26} is null. Again, this can be verified by evaluating tensor \eqref{3.26} on all the possible tensor products of the basis $\{N^i,n^i,E^i_A\}$. The above discussion is summarized by the following
\\
{\Proposition If the separation hypersurface $\Sigma$ is null, eqs. \eqref{3.1tris}, \eqref{3.19}, \eqref{3.22} and \eqref{3.29} are necessary conditions for the quantities \eqref{3.1bb_null} to be distributional solutions of the field equations and the corresponding Riemann tensor to be smooth across $\Sigma$. Instead, eqs. \eqref{3.1tris}, \eqref{3.19}, \eqref{3.22} and \eqref{3.31} are sufficient conditions.}
\\
\\
As above, we can discuss the junction conditions without imposing any regularity on the Riemann tensor $\tilde{R}^h_{\;\,kij}$ in the case of null hypersurface too. In such a circumstance, the junction conditions which must be met come from eqs. \eqref{4.3} and \eqref{3.27bis}. In particular, making use of the projection operator induced by the transverse metric \eqref{transversemetric}, eqs. \eqref{3.27bis} can be decomposed in the equivalent form
\begin{subequations}\label{5.1}
\be\label{5.1.1}
\left[C_{\lambda\lambda}\right]h_{ij} = -2\left(\frac{f''}{f'}\right)^+P^{+h}{}_{ts}n_hh^t_ih^s_j\left[\mathcal{Q}\right] \quad\Rightarrow\quad 
\left[C_{\lambda\lambda}\right]=-\left(\frac{f''}{f'}\right)^+P^{+h}{}_{ts}n_hh^{ts}\left[\mathcal{Q}\right]
\ee
\be\label{5.1.2}
\left[C_{j\lambda}\right]= -2\left(\frac{f''}{f'}\right)^+P^{+h}{}_{ts}n_hN^th^s_j\left[\mathcal{Q}\right], \quad \left[C_{ij}\right]h^{ij}=-2\left(\frac{f''}{f'}\right)^+P^{+h}{}_{ts}n_hN^tN^s\left[\mathcal{Q}\right]
\ee 
\be\label{5.1.3}
\left(\frac{f''}{f'}\right)^+P^{+h}{}_{ts}n_hn^th^s_j\left[\mathcal{Q}\right] = 0, \quad \left(\frac{f''}{f'}\right)^+P^{+h}{}_{ts}n_hn^tn^s\left[\mathcal{Q}\right] = 0, \quad \left(\frac{f''}{f'}\right)^+P^{+h}{}_{ts}n_hN^tn^s\left[\mathcal{Q}\right] = 0
\ee
\end{subequations}
We can again summarize our results by stating the following 
{\Proposition If the separation hypersurface $\Sigma$ is null, eqs. \eqref{4.3}, \eqref{5.1.1}, \eqref{5.1.2} and \eqref{5.1.3} are necessary and sufficient conditions for the quantities \eqref{3.1bb_null} to be distributional solutions of the field equations without any singularity across $\Sigma$.}
\\
\\
Even in the case of null hypersurface, by omitting the regularity of the Riemann tensor, we obtain some junction conditions where Levi--Civita and nonmetricity contributions are coupled (see eqs. \eqref{5.1.1} and \eqref{5.1.2}.

To conclude, we give the junction conditions for the  Lagrangian multipliers appearing in eqs.~\eqref{eq:connection_equation}. According to the usual procedure, the Lagrangian multipliers could be joined by setting
\be\label{3.32}
\lambda_{h}{}^{jip} =  \Theta\lambda_{h}^{+\;jip} + \left(1-\Theta\right)\lambda_{h}^{-\;jip}
\ee
\be\label{3.33}
\lambda_{h}{}^{ij} = \Theta\lambda_{h}^{+\;ij} + \left(1-\Theta\right)\lambda_{h}^{-\;ij}
\ee
Inserting eqs. \eqref{3.32} and \eqref{3.33} into eq. \eqref{eq:connection_equation}, we would get a distributional equation where a singular $\delta$-term proportional to 
\be\label{3.34}
n_p\/\left[\lambda_{h}{}^{jip}\right]
\ee
would appear. The Lagrange multipliers are used to implement a constrained variational problem in an extrinsic way, namely, by working on the entire space of connections without requiring that the deformations of the admissible connections are flat and torsionless. In particular, the constraints on the curvature tensor are non--holonomic (they concern the first--order jet of the connection) and are imposed via a vakonomic variational principle \cite{vakonomic}. As often happens in these cases, the Lagrange multipliers do not appear to be directly connected to any physical or geometrically relevant quantity. 
This, together with the fact that Lagrange multipliers do not enter the Einstein--like equations \eqref{eq:metric_equation}, shows that their junction does not influence the other junction conditions unless a singularity is present in the hypermomentum tensor.  As there is no evidence that singularity exists in standard matter models, we can conclude that the junctions \eqref{3.32} and \eqref{3.33} are not physically relevant.  Clearly, the joining of the Lagrange multipliers might still be required if the regularity of eq. \eqref{eq:connection_equation} across the hypersurface $\Sigma$ were necessary for alternative matter models or ulterior motives. In such a circumstance, the vanishing of the term \eqref{3.34} should be added to the other junction conditions.

\section{Conclusion}\label{Section_IV}
Within the framework of distribution--valued tensors, we discussed the junction conditions for $f({\cal Q})$-gravity. $f({\cal Q})$-gravity is a metric--affine theory where, in addition to the metric, further geometric degrees of freedom appear, given by the nonmetricity tensor. 

As in any metric-affine theory, also in $f({\cal Q})$-gravity, it is necessary to decide on free--fall motions: do the geodesics of the Levi-Civita connection describe them, or are they geometrized by the autoparallel curves of the dynamic connection?

In the first hypothesis, gravity necessarily remains described by the Riemann tensor, i.e., by the curvature of the Levi-Civita connection. In contrast, nonmetricity is responsible for modifying Einstein's equations, possibly accounting for the dark components of the Universe. Clearly, in this scenario, the junction conditions must also include the regularity of the Riemann tensor across the separation hypersurface in addition to the regularity of the joined solutions and the field equations, which are assumed as distribution-valued quantities. In this circumstance, the analysis of the junction conditions is very similar to that which is usually carried out in GR, with results that incorporate and extend those of GR. In fact, on the one hand, we found the same conditions on the jumps of the metric and the extrinsic curvature (or transverse curvature when the separation hypersurface is null) that are valid in GR (they must be zero). On the other hand, we obtained new conditions that involve the jumps of the disformation tensor as well as the second derivative of the function $f(\mathcal Q)$ and the tensor $P^{h}{}_{ij}$ at $\Sigma$, and the jump of the nonmetricity scalar. The relevant point is that the conditions on metric and extrinsic (transverse) curvature on the one hand and those on quantities related to nonmetricity on the other are decoupled.

Nonetheless, junction conditions in $f(Q)$-gravity can also be studied by removing the requirement for regularity of the Riemann tensor. This could correspond to the scenario in which the free-fall motions no longer correspond to the geodesics of the Levi--Civita connection and gravity is no longer related to the Riemann tensor. Therefore, the regularity of the Riemann tensor would no longer be strictly necessary. In this case, we derived junction conditions that differ markedly from those that apply in GR. In particular, this is true for the jump of the extrinsic curvature (transverse curvature for null hypersurfaces), which now is related to the jump of the nonmetricity scalar together with the values of the second derivative $f''$ and the tensor $P^{h}{}_{ij}$ at $\Sigma$.
Moreover, the jumps of the Levi--Civita connection result are related to the jumps of the disformation tensor. In general, by removing the regularity of the Riemann tensor from the junction requirements, we got junction conditions where Levi--Civita contributions are coupled to nonmetricity ones. This represents the most remarkable difference compared to the case where the Riemann tensor is imposed to be regular. However, it is worth noticing that when the function $f(\mathcal Q)$ is linear, or the jump of the nonmetricity scalar is zero, junction conditions are seen to coincide with those of the first analysis, independently of the requirements imposed on the Riemann tensor. The first circumstance is expected because, for $f(\mathcal Q)$ linear, we recover Symmetric Teleparallel Gravity, which is equivalent to GR, at least at the level of Einstein equations. Instead, the condition $[\mathcal Q]=0$ makes a large number of the obtained junction conditions automatically satisfied and brings the analysis back under the regularity assumption for the Riemann tensor.

Junction conditions find their natural application at the astrophysics level, where usually internal and external solutions to given astrophysical objects must be joined. Therefore, the findings of this note can be useful in the study of astrophysical models within the $f(\mathcal Q)$-gravity.

\

{Funding information and acknowledgments}. This work has been carried out in the framework of activities of the INFN Research Project QGSKY.

{Authors individual contributions}. All authors have contributed equally to the computations and the writing of this contribution.

{Conflict of interest}. The authors declare no conflict of interest.

{Data availability}. This manuscript has no data available.

\end{document}